\documentclass{article}
\usepackage{microtype}
\usepackage{graphicx}
\usepackage{subfigure}
\usepackage{booktabs} 

\usepackage{hyperref}

\usepackage[accepted]{icml2023_arxiv}

\usepackage{amsmath}
\usepackage{amssymb}
\usepackage{mathtools}
\usepackage{amsthm}

\usepackage[capitalize,noabbrev]{cleveref}

\theoremstyle{plain}
\newtheorem{theorem}{Theorem}[section]
\newtheorem{prop}[theorem]{Proposition}

\theoremstyle{definition}
\newtheorem{defn}[theorem]{Definition}

\theoremstyle{remark}

\usepackage[textsize=tiny]{todonotes}

\icmltitlerunning{Pseudo-Euclidean Embeddings}

\begin{document}

\twocolumn[
\icmltitle{Pseudo-Euclidean Attract-Repel Embeddings for Undirected Graphs}




\begin{icmlauthorlist}
\icmlauthor{Alexander Peysakhovich}{fair}
\icmlauthor{Anna Klimovskaia Susmelj}{zurich}
\icmlauthor{Leon Bottou}{fair}
\end{icmlauthorlist}

\icmlaffiliation{fair}{Meta AI Research}
\icmlaffiliation{zurich}{ETH Zurich}

\icmlcorrespondingauthor{Alexander Peysakhovich}{alex.peys (at) gmail.com}

\icmlkeywords{Machine Learning, ICML}

\vskip 0.3in
]



\printAffiliationsAndNotice{\icmlEqualContribution} 

\newcommand{\leon}[1]{{\color{purple}L: #1}}
\newcommand{\alex}[1]{{\color{blue}A: #1}}

\newcommand{\fix}{\marginpar{FIX}}
\newcommand{\new}{\marginpar{NEW}}

\def\month{MM}  
\def\year{YYYY} 
\def\openreview{\url{https://openreview.net/forum?id=XXXX}} 


\begin{abstract}
Dot product embeddings take a graph and construct vectors for nodes such that dot products between two vectors give the strength of the edge. Dot products make a strong transitivity assumption, however, many important forces generating graphs in the real world lead to non-transitive relationships. We remove the transitivity assumption by embedding nodes into a pseudo-Euclidean space - giving each node an attract and a repel vector. The inner product between two nodes is defined by taking the dot product in attract vectors and subtracting the dot product in repel vectors. Pseudo-Euclidean embeddings can compress networks efficiently, allow for multiple notions of nearest neighbors each with their own interpretation, and can be `slotted' into existing models such as exponential family embeddings or graph neural networks for better link prediction. 
\end{abstract}

\section{Introduction}

Network analysis is ubiquitous across many disciplines ranging from the natural \citep{jeong2001lethality,barabasi2004network} to the social sciences \citep{granovetter1985economic,easley2010networks,jackson2010social}. In general, graphs are high dimensional objects and can be difficult to work with. Finding easy representations is thus an important problem in applied machine learning. In the symmetric (or undirected) case a workhorse method are node embedding models \citep{lovasz1999geometric,ng2002spectral,perozzi2014deepwalk,tang2015line,grover2016node2vec,athreya2017statistical,lerer2019pytorch}. In these models each node in a network is associated with an embedding (a.k.a latent vector) and dot products between vectors reflect the strength of an edge between two nodes.

We consider situations where a network exhibits a lack of `transitivity' - $A$ is strongly connected to $B$, $B$ is strongly connected to $C$, but $A$ is \textit{not} connected to $C$. Such unclosed triangles are sometimes called `forbidden triads' \citep{granovetter1973strength}. Despite being forbidden, forces such as heterophily or role similarity in real networks give rise to such triads. A major weakness of dot product embeddings is their inability to easily represent networks that contain many such examples \citep{seshadhri2020impossibility}.

Our key contribution is to consider embedding nodes into a pseudo-Euclidean space. Nodes still receive real vector embeddings, however, there is a modified `inner product' where the latent vectors are split into two parts: one on which nodes attract (similar are more likely to connect) and one on which they repel (similar are less likely to connect). The total strength of an edge is modeled as the dot product of the attract sub-vector minus the dot product of the repel sub-vectors. We refer to this as an attract-repel (AR) embedding. We discuss multiplicity of solutions to an AR decomposition and give a give a method for constructing `minimal' AR embeddings from an adjacency matrix using a combination of convex optimization and eigendecomposition. 

We show that AR embeddings have much better representation capability in real-world graphs than Euclidean ones. We then show that AR embeddings can be used to understand the structure of graphs. In social networks, the relative contribution of $A$ vs $R$ components can be used measure heterophily both at the graph and node level. This measurement is fully latent and does not use any label information.

Pseudo-Euclidean space admit multiple notions of ``nearest neighbors.'' Different notions each have interpretable properties in different types of networks. In co-occurrence graphs neighbors in $R$ space map well onto the notion of `substitutability'. In biological co-activation networks different notions of nearest neighbors appear to map well onto the notions of activation and inhibition.

Finally, we focus on the task of link prediction. The AR decomposition can be slotted into the loss function of any model which outputs node embedding vectors. We show that this can lead to increases in model performance for exponential family embeddings \citep{rudolph2016exponential} as well as graph convolutional networks \citep{kipf2016semi} in intransitive graphs.

There is recent interest in using hyperbolic geometry to perform graph embeddings as many graphs of interest have hierarchical structure and Euclidean embeddings have a hard time representing hierarchies \citep{nickel2017poincare,nickel2018learning,liu2019hyperbolic}. We show that hyperbolic space is a manifold in pseudo-Euclidean space and thus AR embeddings are also able to represent hierarchies.

\section{Related Work}
\subsection{VV' vs. VU' Factorizations}
A large literature in graph embeddings considers factorizing the adjacency matrix as $M \sim VV'$ \citep{athreya2017statistical} or as or as $M \sim f (V V')$ for some choices of $f$ \citep{hoff2002latent,rudolph2016exponential}. Other methods work indirectly on the adjacency matrix by constructing the embeddings from co-occurrences in random walks \citep{perozzi2014deepwalk,grover2016node2vec,tang2015line}.

Recently \citet{seshadhri2020impossibility} show that Euclidean dot product models cannot reproduce the distribution of triangles and degrees in real world social networks, no matter what algorithm is used to construct the embeddings. \citet{chanpuriya2020node} respond and argue for factorizing adjacency matrices as $M \sim VU'$. Our results show that the $M \sim VU'$ formalization is `too general' for undirected graphs since symmetry implies $u_i v_j = u_j v_i$, so $VU'$ can be written in a pseudo-Euclidean form $M \sim AA' - RR'$.



\subsection{Non-Metric Visualization}
 \citet{van2012visualizing} considers extending t-SNE \citep{van2008visualizing} to intransitive similarity matrices by embedding objects in multiple t-SNE maps simultaneously - i.e. in multiple vector spaces each of dimension $2$ (because of the interest in visualization). Constructing visualization techniques that take advantage of the unique geometry of pseudo-Euclidean space is an extremely interesting future direction.

\subsection{The Eigenmodel}
The `Eigenmodel' is the closest work to our own \citep{hoff2007modeling}. It decomposes the adjacency matrix into a learned $VDV'$ where $D$ is a diagonal matrix. Our work builds upon this in several ways. First, we formalize much of the intuition in that paper. Second, we give a method for guaranteed computation of the `simplest' AR decomposition. Third, we study properties of the model beyond just better out of sample fit (e.g. interpretation of `neighbors'). 

\subsection{Directed Graph Embeddings}
Our work intersects a large literature on \textit{directed} graphs. The ComplEx approach approximates knowledge graphs as $A \sim EVE'$ where $V$ is the diagonal matrix of eigenvalues which can take complex values \citep{trouillon2016complex}. When the matrix is symmetric $V$ is guaranteed real so it is very close to the Eigenmodel above. \citet{sim2021directed} considers the intransitivity problem in undirected graphs and performs a pseudo-Riemannian embedding to deal with it. The symmetry of undirected graphs gives us the additional structure that allows us to use off-the-shelf optimization techniques as well as good interpretability properties.

\section{Dot Product Embeddings}
We consider the general problem of embedding a weighted undirected graph $G = (N, E)$. Nodes $N$ are generically indexed by $i,j$, edges $e_{ij}$, with $e_{ij} = e_{ji}$. We ignore self-edges in the graph, so $e_{ii}$ is not defined.

We consider embeddings into $\mathbb{R}^{D}$ endowed with the dot product $x \cdot y = \sum_{j=1}^d x_j y_j.$ We refer to a set of vectors, one for each node, $\mathcal{V} = \lbrace v_1, \dots, v_N \rbrace \subset \mathbb{R}^D $ with the dot product as an Euclidean embedding of the graph. Let say that $D$ is the dimension of the embedding.

\begin{defn}
We say that a Euclidean embedding $\mathcal{V}$ \textbf{represents the graph} if for all $i \neq j$ we have $$v_i \cdot v_j = e_{ij}.$$
\end{defn}

Our interest will be the dimension of the embedding $\mathcal{V}$. 

\begin{prop}\label{thm:euc_exists}
Let $G$ be an arbitrary graph. There exists an infinitely sized family of dot product embeddings $\lbrace \mathcal{V} \rbrace$ that represents $G$ 
\end{prop}



We leave the proof to the Appendix. However, while the Proposition shows that Euclidean embeddings always exist, we will now see that this embedding may not be low dimensional even when the underlying graph is `simple'.

\begin{center}
\begin{tabular}{| p{3cm} | p{3cm} |} 
 \hline
  \includegraphics[scale=.5]{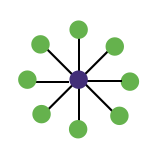} &  \includegraphics[scale=.5]{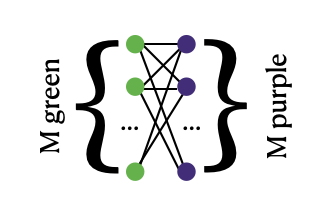} \\ 
  \hline
 Star & M-Bipartite \\ 
 \hline
\end{tabular}
\end{center}

Consider the graphs above with edge weights $e_{ij} \in \lbrace 0, 1 \rbrace$ with colors only added for visualization purposes. For the dot product to represent any star graph any two green nodes must have $v_i \cdot v_j = 0$, but this means they must be orthogonal. Thus, in the star graphs, we need at least as many dimensions as peripheral nodes. The M-bipartite graph is another example: this graph has a very simple structure but it is maximally intransitive (if $i$ is linked to $j$, it is never linked to any neighbor of $j$). We will now formally show that any embedding of this graph does not really compress it. We will see later that both of these graphs have $2$ dimensional pseudo-Euclidean embeddings for any choice of $M$.

\begin{prop}\label{thm:no_euclid}
If $\mathcal{V}$ is a Euclidean embedding that represents the $M$-bipartite graph then it has dimension $\geq 2M-1$.
\end{prop}

Again, we relegate the proof to the Appendix.


\section{Pseudo-Euclidean Embeddings and Attract-Repel}\label{AR_construct}
Let us consider the counterexample above more. In the $M$-bipartite graph all green nodes have the same connectivity pattern, so for maximum compression we would like to give them the same representation, however, such a construction means they will have high dot product with each other. So, such an efficient encoding is impossible in the dot product model. We will now work with pseudo-Euclidean space where vectors can have inner product $0$ with themselves and where the triangle inequality does not apply, thus allowing for efficient coding of the $M$-bipartite and other intransitive graphs.

Given $\mathbb{R}^M$ we endow it with $\cdot^k$ with $k>0$ which is a symmetric bilinear form (which is not technically an inner product but we will refer to as one) defined as $$v_i \cdot^k v_j = \sum_{p=1}^{M-k} v_{ip} v_{jp} - \sum_{q=M-k+1}^{M} v_{iq} v_{jq}.$$ 

Setting $k=0$ we have standard Euclidean space. The special case of $M=4$ and $k=1$ is known as Minkowski spacetime \citep{naber2012geometry}. We consider the general $(M, k)$ case here.

For simplicity of notation, instead of considering a single vector $v_i$ per node, we split the vectors into $a_i$ and $r_i$ so that we can write the score as $a_i \cdot a_j - r_i \cdot r_j$ where $\cdot$ is the standard dot product. We refer to this as an attract-repel (AR) embedding. We can still define an embedding as representing a graph:

\begin{defn}
We say that \textbf{a AR embedding represents $G$} if for any two $i, j$ with $i \neq j$ we have that $$e_{ij} = a_i \cdot a_j - r_i \cdot r_j.$$
\end{defn}

Both the star and $M$-bipartite graphs have high dimensional Euclidean-only representations but a simple AR decomposition where $a_{i} = 1$ for all $i$ and $r_i = 1$ if $i$ is green and $r_i = -1$ if $i$ is purple. We know from the prior section that a Euclidean (and thus an AR embedding with R empty) always exists so the AR problem is over-parametrized in a non-trivial way.

\subsection{Minimal Pseudo-Euclidean Embeddings}
We now discuss how `minimal' AR embeddings can be found. Let $\mathcal{AR}$ be the set of all AR embeddings representing a graph $G$ and let $A$ and $R$ be the stacked embedding vectors for each node. We will look for the solution with the smallest Frobenius norm, a solution to $\min_{(A,R) \in \mathcal{AR}} ||A||^2_{F} + ||R||^2_{F}.$ We first show that this solution can be found using convex optimization and eigendecomposition which gives better guarantees than a local search.

Start with a graph $G$ and consider the adjacency matrix of $G$. Recall that we do not look at self-edges in our graph, therefore any choice of diagonal for the matrix makes it a valid adjacency matrix for all $i \neq j$. Let $D$ be an $M$-dimensional vector and denote by $M_{D}$ as the matrix that is the adjacency matrix for $G$ on the off-diagonal and has arbitrary diagonal $D$. 

We now show that finding the simplest $AR$ decomposition is strongly related to finding an appropriate choice for $D$. In particular, we choose $D$ to minimize the nuclear norm of $M_{D}$  \citep{candes2009exact}. More formally:

\begin{prop}\label{thm:nuc_norm}
Let $A,R$ be a solution to $\min_{(A,R) \in \mathcal{AR}} ||A||^2_{F} + ||R||^2_{F}.$ Let $M_{D}$ be the solution to $\min_{D} || M_{D} ||_{*}$. Then $M_{D} = A'A - R'R.$
\end{prop}

We leave the proof to the Appendix as it uses standard techniques from the literature. However, we use this equivalence to construct the lowest norm AR embedding:

\begin{algorithm}[h!]
   \caption{Construct Minimal AR Decomposition}
   \label{alg:example}
\begin{algorithmic}
   \STATE Solve the convex problem: $$\min_{\hat{M}} || \hat{M} ||_{*}
\text{ s.t. }  \hat{M}_{ij} = e_{ij} \forall i \neq j$$
   \STATE Compute the eigendecomposition of $\hat{M} = Q'DQ.$
   \IF{low rank is desired}
   \STATE Truncate the $n-k$ smallest in absolute value eigenvalues to $0$
   \ENDIF 
   \STATE Let $D^{-}$ be the strictly negative eigenvalues 
   \STATE $D^{+}$ be the strictly positive ones
   \STATE Let $Q^{-}$ correspond to the eigenvectors with negative eigenvalues and $Q^{+}$ be the eigenvectors with positive eigenvalues.
   \STATE Set $A = Q^{+} \sqrt{D^{+}}$ and set $R = Q^{-} \sqrt{-D^{-}}$
   \STATE Rows of $A$ are $a_i$, rows of $R$ are $r_i$
\end{algorithmic}
\end{algorithm}


For relatively small matrices, we can solve nuclear norm minimization directly. However, it scales poorly with matrix size. A popular solution for medium size approximate solutions to the nuclear norm is SVT (\citet{cai2010singular}). We use code implemented in the $R$ package \textit{filling} \citep{filling}. 

Often we are willing to take a lossy compression of our data - i.e. a low rank representation. To select the `natural' rank we use generalized Gabriel bi-cross-validation (BCV) \citep{owen2009bi}. In BCV the row and column indices are split into folds, one fold of the matrix, is held out while the rest of the matrix is used to fit a low-rank factorization. The estimated `natural' rank of the matrix is the one which minimizes average held out loss. We point the readers to the exposition in \cite{owen2009bi} which discusses the guarantees of BCV as well as advantages of this method over many other choices. However, any method for rank selection for can be used in the procedure above. 

For cases where SVT cannot be applied, we will use gradient descent methods and add an explicit regularizer on $||A||, ||R||$ when we deal with link prediction.


\section{Hierarchical or `Tree-Like' Graphs and AR} 
A recent literature focuses on another failure point of Euclidean embeddings: they are poor at representing hiearchical graphs. A proposed solution to this problem is instead embedding graphs into hyperbolic space which has better representation capacity for such graphs \citep{nickel2017poincare}. Note that hyperbolic models continue to be metric, yet we will see there is a deep relationship between pseudo-Euclidean embeddings and hyperbolic ones.

We follow the exposition in \citet{nickel2017poincare} to introduce this model. The Poincare model of $d$-dimensional hyperbolic space is given by the open ball $\mathcal{B}^d = \lbrace x \in \mathbb{R}^d | || x || < 1 \rbrace$ endowed with the metric tensor $g_{x} = (\dfrac{2}{1 - || x ||^2}) g^E$ where $g^E$ is the Euclidean metric tensor.

The distance between any two points in the poincare model is given by $$d(x,y) = \text{arccosh} (1 + 2 \dfrac{||x - y||^2}{(1-||x||^2)(1-||y||^2)}).$$

While so far we have focused on exact representations that require $e_{ij} = v_i \cdot v_j$, the experimental measures in \citet{nickel2017poincare} use a slightly different criterion, easier to apply in unweighted ($e_{ij} \in \lbrace 1, 0 \rbrace$) graphs:

\begin{defn}
A set of vectors $\mathcal{V}$ with metric $d$ order-represents an unweighted graph if for any $i,j$ with $e_{ij} = 1$ and $k$ with $e_{ik}=0$ we have $d(i, j) < d(i,k).$ In other words, for any node any other nodes it is connected to are closer to it in embedding space than any non-connected nodes.
\end{defn}

For our comparison purposes we will work with this weaker requirement. Adapting the definition to the AR product means simply replacing the $d(i,\cdot)$ with the AR product and reversing the inequality (since in distance more similar is smaller but in inner product more similar is larger). Given these definitions we can now show the following result:

\begin{prop}\label{thm:hype_equiv}
Let $G$ be a graph that is order-represented by a hyperbolic embedding $\mathcal{V}$ of dimension $d$. Then there exists a $d+1$ dimensional AR embedding with $R$ dimension $1$ that also order-represents the graph.
\end{prop}

The proof of this Proposition is relatively straighforward and uses fact that there is a diffeomorphic model of hyperbolic space called the Lorenz model \citep{nickel2018learning}. We relegate it to the Appendix.

\section{``Neighbors'' in Pseudo-Euclidean Space}\label{neighbors}

One of the most common uses of embeddings in practice is clustering or nearest neighbor lookup. Formally, the problem is: given a query vector $q$ and a database of node embeddings for all nodes $j\neq q$ we want to find the nearest neighbor of $q$.\footnote{ This is used in many real world machine learning pipelines under the name of `vector databases' \citep{raghavan1986critical} with recent interest in constructing fast, large-scale, nearest neighbor lookup libraries \citep{johnson2019billion}.} Normal Euclidean embeddings have a single notion of nearest neighbor given by $\text{argmax}_{j} v_q \cdot v_j$.

In this section we will show 1) there are at least 4 interesting notions of `neighbor' in pseudo-Euclidean space, each has a different interpretation, 2) in practice we can still use Euclidean nearest neighbor libraries by appropriately formatting the query vector. 

First, we write the database of vectors $\mathcal{D}$ to be searched as the concatenation of vectors $[a_j, r_j]$ for each node $j$. We can represent the query $q$ as the concatenation $[a_q, -r_q]$. If we compute dot product neighbors of $q$ we get back nodes with high values of $a_q \cdot a_j - r_q \cdot r_j$. This corresponds to nodes $j$ with high $e_{qj}$ in the original graph. We will refer to this as `first order' similarity of nodes and denote it by $F(q, j).$

We can also represent $q$ as $[a_q, r_q].$ Finding standard dot product nearest neighbors in this case results in nodes $j$ with high values of $a_q \cdot a_j + r_q \cdot r_j.$ If $q$ and $j$ have a very high value of this product, it means that for any other node $k$, $e_{ij}$ and $e_{ik}$ are very close. Thus, this returns nodes with similar neighborhoods to $q$, which is sometimes called `second order' similarity \citep{tang2015line}, we denote by $S(q, j)$.

We consider the difference $S(q,j) - F(q,j)$. Given a query node $q$ a node $j$ scores high on this composite metric if it has the same neighbors but is not connected to $q$. In other words, if $q$ and $j$ are part of many forbidden triads. We will later see that this corresponds to `substitute' pairs in co-ocurrence graphs. Replacing $F, S$ by their definitions gives $a_q \cdot a_j + r_q \cdot r_j - a_q \cdot a_j - r_q \cdot r_j = 2 r_q \cdot r_j.$ Since distances here are dimensionless we can replace this with $r_q \cdot r_j$ - i.e. the dot product nearest neighbor in $R$ space or a lookup using the $[0, r_q]$ as the query vector.

The final notion we consider are nodes $j$ that score high on $S(q,j) + F(q,j)$. These nodes are nodes which have similar neighborhoods \textit{and} are strongly connected to each other. In other words, these are nodes that are part of many triangles with $q$. The same argument as the paragraph above gives that these are nodes with high similarity in $A$ space, $a_q \cdot a_j.$ This corresponds to standard nearest neighbors in $A$ space or lookups using the query vector $[a_q, 0]$.

\begin{figure}
\centering
\begin{small}

\begin{tabular}{p{2cm} p{5cm}}
  \hline
Proximity & Interpretation \\ 
  \hline
$a_i \cdot a_j - r_i \cdot r_j$ & Measures whether $i,j$ are directly connected\\ 
\hline

 $a_i \cdot a_j + r_i \cdot r_j$ & Measures whether $i,j$ connected to same other nodes \\ 
\hline
  $r_i \cdot r_j$ & Measures whether $i,j$ are part of forbidden triads - i.e. connected to similar others but not to each other - useful in finding `substitute' nodes  \\ 
\hline
  $a_i \cdot a_j$ & Measures whether $i,j$ are part of many triangles - connected to same other nodes and to each other \\
  
   \hline

\end{tabular}
   \end{small}
   \caption{While Euclidean space has a single notion of proximity, pseudo-Euclidean embeddings admit multiple notions of proximity between nodes. Here we summarize four different ways of computing node proximity and their interpretation in practice. }
\end{figure}

\section{Empirical Evaluation}
We now turn to an empirical evaluation of pseudo-Euclidean embeddings. We will evaluate representational capacity (experiments 1, 2), ability to learn about graphs from different notions of neighbors (experiments 3A, 3B, 4A, 4B, 5), and finally generalization capacity when slotted into commonly used link prediction models (experiment 6).   


\subsection{Experiment 1: AR vs Euclidean on Social Graphs}
Our theoretical results show that AR embeddings require fewer, and sometimes drastically fewer, dimensions to perfectly represent the same graph as a dot product embedding. We first ask: does this hold for approximate representation?

We consider the \href{https://snap.stanford.edu/data/deezer_ego_nets.html}{anonymized ego-networks} (an ego network takes a focal ego, takes all of their friends, and maps the friendships between them) of $627$ users of a music social network \citep{karateclub}. We consider users with at least $50$ friends (mean ego network size = 81.6). 

\begin{figure}
\begin{center}
\includegraphics[scale=.5]{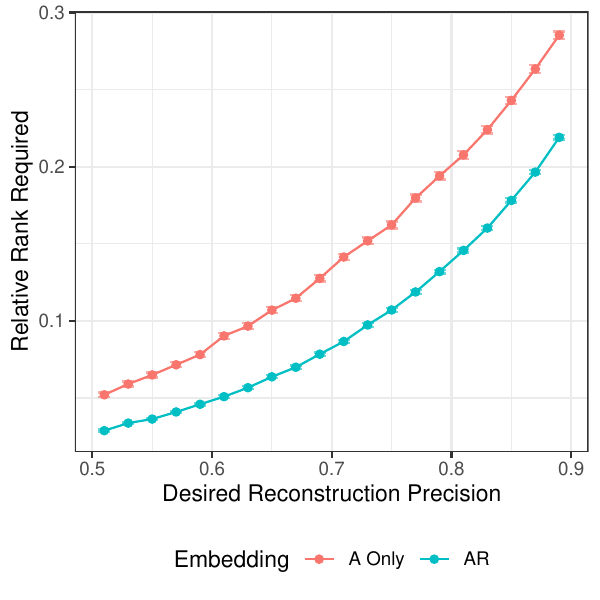}
\caption{AR embeddings are much more efficient at compressing social networks than dot product embeddings. Error bars (very small) reflect standard errors.}\label{deez_het}
\end{center}
\end{figure}

We construct minimal AR embeddings as described above. Letting $e_{ij}$ be the true edges and $\hat{e}_{ij}$ be the model estimated edges we first consider the variance explained in $e$ by $\hat{e}$ (reconstruction precision). We consider what dimension of embedding is required to achieve a given reconstruction quality across our $627$ networks in Figure \ref{deez_het}. The standard dot product requires a $\sim 50 \%$ higher dimensionality to recover the network with the same fidelity as the AR decomposition.

\subsection{Experiment 2: Representing Hierarchical Graphs}

Theory guarantees that if hyperbolic embeddings can represent a graph, an AR representation also exists. However, just because a solution exists doesn't mean that gradient descent - the most common way that embeddings are trained in practice - can actually find it. This is what we study here using an experiment similar to \citet{nickel2017poincare} looking at the \textit{transitive closure} of the mammal subtree of WordNet \citep{miller1995wordnet}.

For the hyperbolic embeddings we use code directly from \href{https://github.com/facebookresearch/poincare-embeddings}{the paper repository}.

A standard method for embedding unweighted (i.e. $e_{ij} \in \lbrace 0, 1 \rbrace$) graphs into Euclidean space is using an exponential family link function \citep{hoff2002latent,rudolph2016exponential} which we will refer to as a \textbf{logistic node embedding} (LNE). In the LNE we model $p(e_{ij} = 1) = \sigma (v_i \cdot v_j)$ where $\sigma$ is the sigmoid function. Another way to think about LNE is that it is a Euclidean embedding of the matrix of logits (rather than the original binary edges). 

LNE is trained with binary cross entropy loss on the binary edge labels. Since the graphs are sparse, we need to use negative sampling. We use the strategy introduced in \citet{lerer2019pytorch} - for every positive sample $e_{ij}$ we consider two `corruptions' $e_{ik}$ where $k$ is not a true neighbor of $i$ in the graph. In one we take $k$ sampled uniformly from the set of non-neighbors, in the other we take $k$ sampled proportional to its degree. This means that more common nodes are represented more highly, but for graphs with fat tailed degree distributions they do not completely dominate the set of negative samples. 

We consider the AR extension of this model (LNE-AR) by using the AR product instead of the dot product giving $p(e_{ij} = 1) = \sigma (a_i \cdot a_j - r_i \cdot r_j).$ In this experiment we use a single $R$ dimension as suggested by our theorem, we refer to this as LNE-AR1.

\begin{figure}[h!]
    \centering
    \includegraphics[scale=.4]{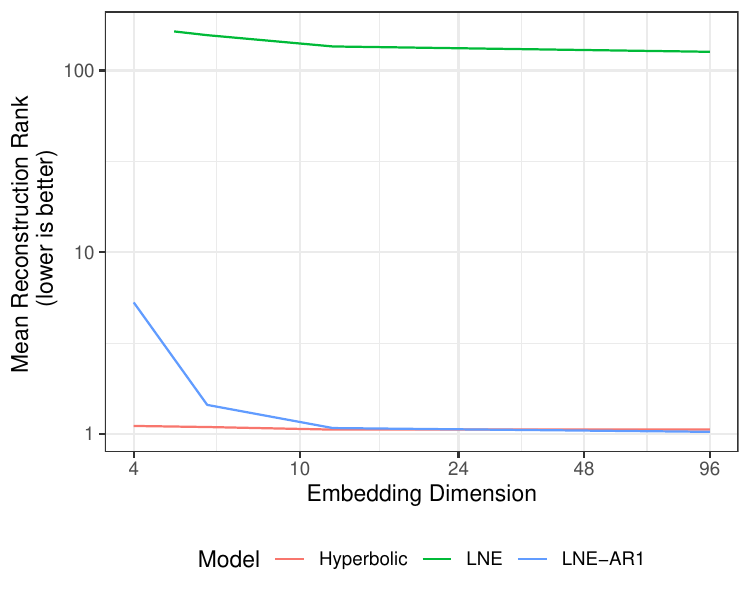}
    \caption{Pseudo-Euclidean with a single $R$ dimension (LNE-AR1) embeddings are able to represent hierarchical relations while Euclidean (LNE) embeddings are not.}
    \label{hype_fig}
\end{figure}

As in  \citet{nickel2017poincare} we use mean reconstruction rank which takes every real edge $(i, j)$ and all corresponding negative edges $(i,k)$ where $e_{ik} = 0$ and asks how many $k$ rank above $j$ in terms of dot product/distance. In essence, we ask: are true neighbors of $i$ closer in embedding space than non-neighbors?


Figure \ref{hype_fig} shows our results. LNE fails completely to represent the graph even in high dimensions. We also see that the optimization does not quite reproduce the bound of our theorem: a $4$ dimensional hyperbolic embedding represents the graph perfectly but a $5$ dimensional LNE-AR1 still is slightly behind in terms of representation accuracy. This small discrepancy is likely due to many factors including the fact that while the hyperbolic code optimizes directly for the contrastive loss, the LNE optimizes for the classification loss and that the negative sampling and optimization procedures are off the shelf and not tuned. Nevertheless, the experiment shows that AR embeddings can work well in real world graphs that exhibit both intransitivity and hierarchy.

\subsection{Experiment 3: Measuring Homophily and Heterophily using AR}
We now begin to ask whether there are gains from using AR embeddings from an interpretability standpoint. That is, can we learn interesting things about the graph directly from the embeddings?

In the case of social networks, there is a straightforward interpretation of the AR decomposition. There are latent attributes $A$ on which birds of a feather flock together (i.e. the homophily in the network) and there are latent attributes $R$ where opposites attract (the heterophily in the network).

In the AR decomposition we can see how much of a network is explained by the $R$ component by looking at $\dfrac{||R||^2_{F}}{||A||^2_{F} + ||R||^2_{F}}$. We call this the $R$-fraction of the network. We can think of this as a latent measure of heterophily. 

\begin{table}
\centering
\begin{small}

\begin{tabular}{p{2cm} p{2cm} p{3cm}}
  \hline
Network & R-Fraction & Node Assortativity \\ 
  \hline
Wisconsin & .59 & .15 \\ 

 Texas & .66 & 0.05 \\ 

  Cornell & .68 & 0.11 \\ 

  Citeseer & .81 & 0.72 \\ 
  Cora & .81 & 0.82 \\ 

  EU & .87 & 0.46 \\ 
  
   \hline

\end{tabular}
   \end{small}

\caption{Graph level $R$-fraction predicts node assortativity across graphs. Node assortativity is a commonly used heuristic for identifying a graph as homophilous or heterophilous.}\label{global_AR}
\end{table}

To evaluate whether $R$-fraction is a useful measure we consider symmetric versions of standard datasets from the graph literature: Cora \citep{mccallum2000automating}, Citeseer \citep{giles1998citeseer}, WebKB-Wisconsin, WebKB-Cornell, WebKB-Texas \citep{craven1998learning}. We also include the EU e-mail dataset \citep{leskovec2007graph,yin2017local}. Each dataset has a label for each node and thus we can use node assortatitivity (fraction of neighbors sharing focal node's label) as an observed proxy for heterophily in the dataset as is defined in the recent literature on heterophilic GNNs  \citep{zhu2021graph,zheng2022graph}. 

We begin by seeing whether the $R$-fraction of a network predicts its label assortativity. We use the algorithm outlined in Section \ref{AR_construct} to build low rank AR representations. In Figure \ref{global_AR} we see that the $R$ fraction of these low rank representations indeed predicts label assortativity at the network level.

\subsection{Experiment 3B: Local Measures of Heterophily}

\begin{figure}
\begin{center}
\includegraphics[scale=.45]{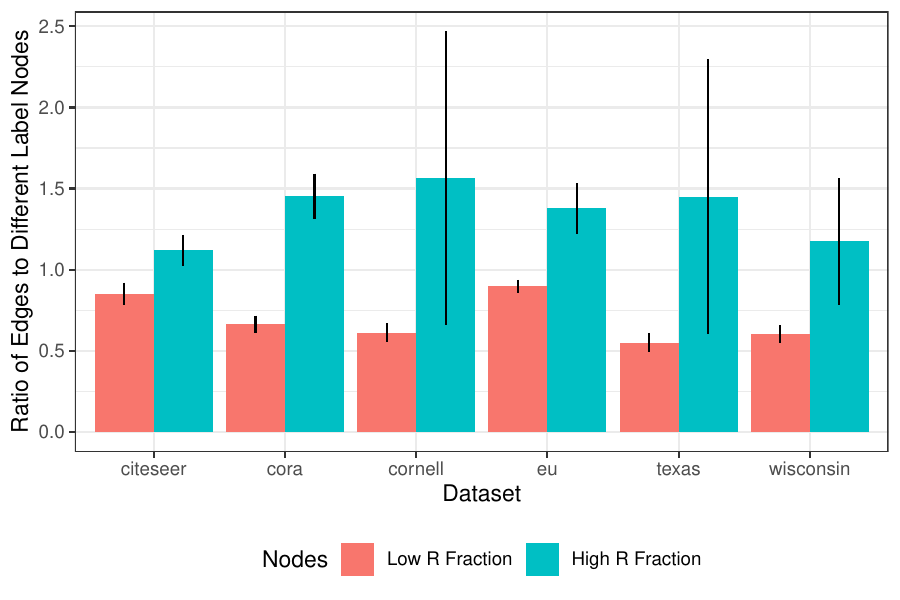}
\caption{Nodes with higher $R$-fractions than the median in the graph have more links to nodes with different labels. Error bars reflect standard errors computed at the node level.}\label{local_AR}
\end{center}
\end{figure}

The analysis above looks at the graph as a whole, but the same idea can be applied to each node. Given an AR embedding, we can compute the $R$-fraction for each node individually $\dfrac{||r_i||}{||r_i|| + ||a_i||}.$ We ask whether individuals that have relatively high $R$-fractions are more likely than low-$R$ fraction nodes to be connected across labels. In Figure \ref{local_AR} we take each graph, consider nodes with at least $5$ neighbors, and median split these nodes by $R$-fraction. We then look at the number of edges they have to nodes with labels not the same as their own. To make comparisons across networks with very different degree distributions we normalize by the average number of edges that a node has to other nodes of different labels - in other words, we ask: do nodes whose $R$-fraction is above the median have more edges to nodes with different labels than average? In Figure \ref{local_AR} see that the answer is yes. Again, label data is \textit{not used} at all during $AR$ embedding construction, so we are reading per-node `heterophily' purely from the graph. 

\subsection{Experiment 4: Finding Substitutes using AR}
There is recent interest in using embedding techniques to find substitutable products \citep{ruiz2020shopper}. Substitutes in this case are defined as products which fulfill the same need - or, in the case of co-purchase graphs, are purchased with the same items but rarely together. For example, both Pepsi and Coke may be purchased with Hamburgers and Fries, but a purchase which contains Pepsi usually does not also contain Coke. Section \ref{neighbors} shows that, in theory, we can find such pairs by looking at neighbors in $R$ space. We now ask whether this yields meaningful substitutes in practice.

\subsubsection{Experiment 4A: Roles on Teams}
We begin by looking at data from the online game \href{https://www.dota2.com/home}{DotA2}. In this game individuals are placed in a team of $5$, each individual chooses one of $115$ (as the time of this analysis) `heroes', and the team competes against another team. As with many team sports, there are different roles on a team that need to be covered and so real world teams are unlikely to include multiple copies of the same role. Heroes in DotA are different and specialized, each able to play only a subset of roles. 

We use a publicly available \href{https://www.kaggle.com/c/mlcourse-dota2-win-prediction/overview}{Kaggle dataset} of $39,675$ DotA matches. From this data we construct a co-occurrence matrix for the heroes. Letting $c_{ij}$ be the co-occurrence between $i$ and $j$. Because the co-occurences are extremely right skewed, we consider the matrix of $\text{log}(c_{ij}+1)$ though qualitatively all our results go through using the raw co-occurrence counts as well. We take the low rank (k=10) exact $AR$ decomposition of this co-occurrence matrix.


\begin{figure}
\begin{center}
\includegraphics[scale=.6]{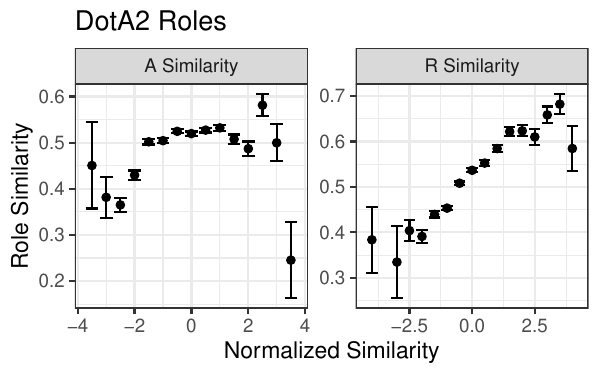}
\caption{In our DotA data we see that similarity in $A$ vectors does not predict similarity in roles very well but similarity in $R$ vectors (produced without knowing roles) does. Error bars reflect standard errors computed at the bin level.}\label{dota_fig}
\end{center}
\end{figure}

For each possible hero, we take the list of `official roles' the hero can play from the \href{https://dota2.fandom.com/wiki/Role}{the DotA wiki}. We construct a vector for each hero where a $0$ in a dimension indicates that the hero cannot play that role and $1$ indicates they can. We then ask whether, given two heroes, the similarity in these `true role vectors' is predicted by their similarity in $A$ or $R$ space. Since similar roles are substitutes, we should expect to see $R$ but not $A$ similarity to be related to role similarity, which is precisely what we see in Figure \ref{dota_fig}. Again, the embeddings $A$ and $R$ do not use any role labels in their construction, only co-occurrence counts.

\subsubsection{Experiment 4B: Substitutes in Ingredients}
\begin{figure}
\centering
\begin{tiny}

\begin{tabular}{p{2cm} p{2cm} p{1cm}}
  \hline
target & substitute & score \\ 
  \hline
baking mix & bisquick & 0.78 \\ 

  baking powder & baking soda & 0.85 \\ 

  beer & apple juice & 0.40 \\ 

  brown sugar & sugar & 0.74 \\ 
  buttermilk & skim milk & 0.52 \\ 

  chicken broth & vegetable broth & 0.63 \\ 

  lemon & fresh lemon juice & 0.76 \\ 
  
  onion & scallion & 0.71 \\ 

  orange juice & honey & 0.61 \\ 

  parmesan cheese & mozzarella & 0.64 \\ 

  parsley & dried parsley & 0.59 \\ 
  pecan & walnut & 0.85 \\ 
  pecan & sliced almond & 0.65 \\ 
  red wine & dry white wine & 0.68 \\ 

  unsalted butter & margarine & 0.67 \\ 

  unswtd chocolate & baking cocoa & 0.74 \\ 

  vegetable oil & canola oil & 0.88 \\ 

  vinegar & cider vinegar & 0.89 \\ 

  yogurt & greek yogurt & 0.70 \\ 
   \hline

\end{tabular}
   \end{tiny}

\caption{Substitutes for various focal ingredients found by looking at $R$ neighbors.}\label{ingsub}
\end{figure} 

We now look at a different substitute task. We use a \href{https://www.kaggle.com/shuyangli94/food-com-recipes-and-user-interactions}{dataset} of $180,000+$ cooking recipes \cite{majumder2019generating}. We construct the log co-occurrence matrix of the $1000$ most common ingredients in these recipes. We compute the exact low rank (k=125) AR decomposition of this matrix.

We then look at some \href{https://www.allrecipes.com/article/common-ingredient-substitutions/}{commonly substituted cooking ingredients}. We restrict to focal ingredients that appear in the 1000 most commonly used ingredients and have exact 1-1 substitutes rather than mixtures of items. In Table \ref{ingsub} we take some focal ingredients and show their nearest $R$ neighbors using the cosine similarity. We use cosine similarity as the length of an $R$ or $A$ vector encodes a node's commonality. We include only the top neighbor for space here, in the Appendix we include an expanded version of the table including the top $3$ suggested substitutes per target ingredient. We see that using $R$-similarity as a substitutability metric seems to yield qualitatively good results in this dataset.

\subsection{Experiment 5: Inhibition and Activation in Biological Networks}

Systems biology is a field focusing on study of interactions between genes or proteins. Deterministic or stochastic dynamical systems are usually used to model these interactions. However, the topology of the governing equations is quite often partially or fully unknown. We ask whether AR embeddings can help researchers recover information about the directed graph of structural equations governing interactions from observed co-occurrence relationships.

Most real gene regulatory networks are poorly understood, so simulations of a gene regulatory networks are often used. In our example we use a commonly used, simplified model of hematopoietic stem cell differentiation \cite{krumsiek2011hierarchical}.  This network consists of 11 transcription factors with 28 directed regulatory interactions between them. Some  exhibit activation relationships ($x$ makes $y$ more likely) and some of which exhibit inhibition ($x$ makes $y$ less likely), see the Appendix \ref{appendix_exp5} for a full description. We sampled snapshots of expressions from the system. From these snapshots we construct the co-occurrence matrix of transcription factors. We compute the AR decomposition of this matrix using the same methodology as the experiments above (rank = 8). 

\begin{figure}[h!]
    \centering
    \includegraphics[scale=.35]{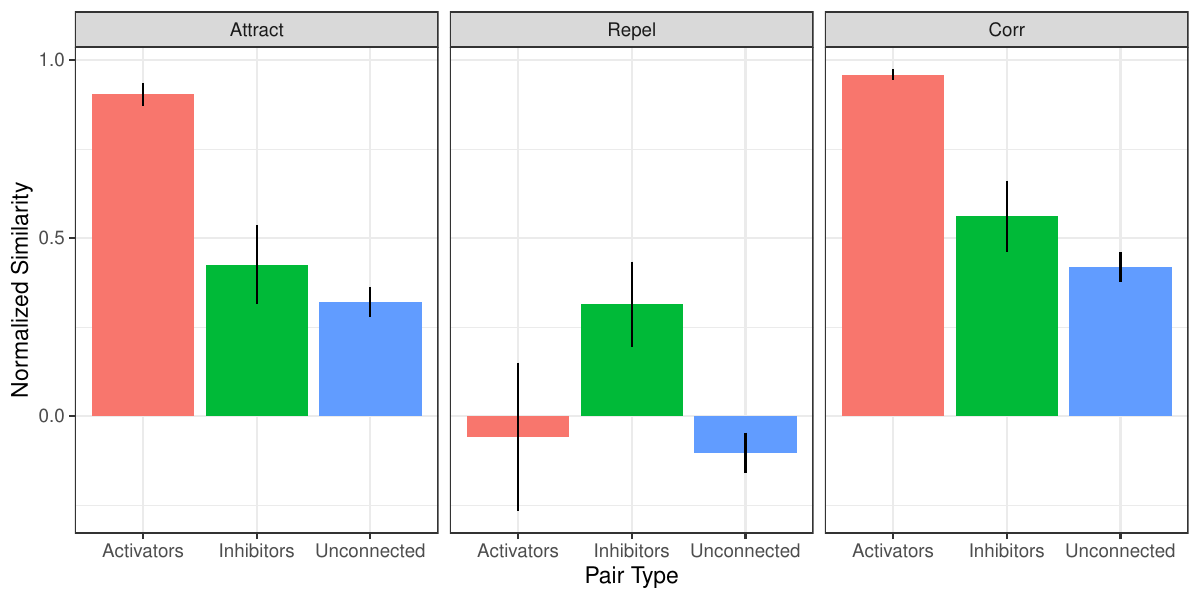}
    \caption{Activators are closer in $A$ space than average while inhibitors are close in $R$ space. Euclidean embeddings are less clear. Error bars reflect standard errors.}
    \label{fig:krum2}
\end{figure}

We then compare the similarity in $A$ and $R$ components across inhibitor and activator pairs. Importantly, while inhibition/activation are directed relationships, we only observe undirected correlations. In Figure \ref{fig:krum2} we see that activators are closer in $A$ space while inhibitors are closer in $R$ space. Looking at the Euclidean embedding of the correlation between two nodes does not display as clean of a pattern. See the Appendix \ref{appendix_exp5} for another analysis visualizing the embeddings.

\subsection{Experiment 6: Pseudo-Euclidean Embeddings in Other Models}
In this section, we move from the problem of reconstruction (asking how well models can express certain data) and consider the task of link prediction - in other words, generalization to unseen edges. 


The first models we consider are the LNE/LNE-AR from above. An increasingly popular method for dealing with graph data are graph neural networks (GNN) \citep{zhou2020graph}. Typically GNNs are used in graphs where nodes have feature vectors to do inductive classification tasks, however since GNNs construct a vector for each node (call this $v_i^{GNN}$) we can also use these vectors for link prediction tasks \citep{zhang2018link} by using $\sigma (v^{GNN}_i \cdot v^{GNN}_j)$ as our edge probabilities. We can convert the GNN to do the AR embeddings simply by splitting the $v^{GNN}$ vectors and using the LNE-AR formula for edge probabilities.

Our goal is to investigate bonuses provided by AR rather than trying to achieve perfect state of the art performance. Thus, we focus on the simple graph convolutional network (GCN) \citep{kipf2016semi}. We train LNE, LNE-AR, GCN, and GCN-AR models on Cora, Citeseer, Wisconsin, Texas, Cornell which all have feature vectors for all nodes. We do not use the EU data or the ego network data as they do have node features and so are less interesting from the GCN standpoint. 

We split each dataset into $80/10/10$ train/validation/test. We adapt the hyperparameters, training, evaluation, and error bar construction code from \citet{chami2019hyperbolic} for all of our experiments. See Appendix \ref{appendix_exp6} for more details.

\begin{figure}[h!]
    \centering
    \includegraphics[scale=.5]{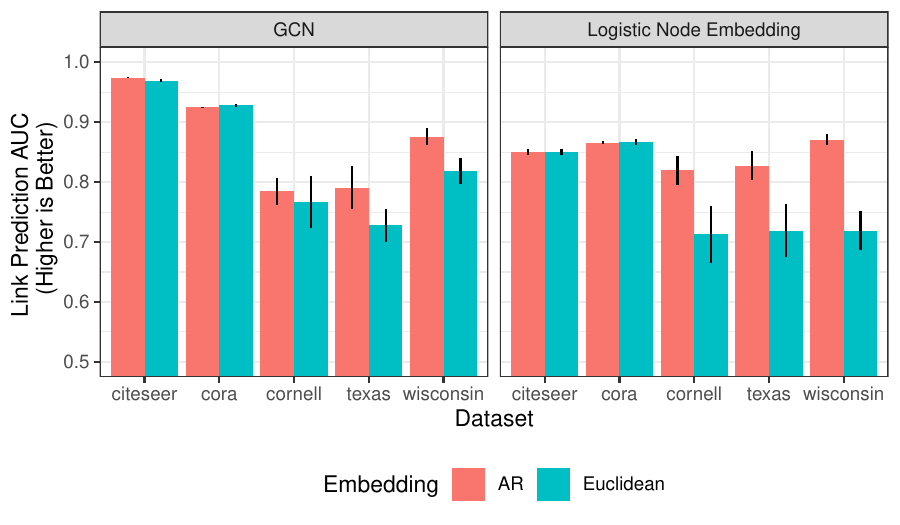}
    \caption{Pseudo-Euclidean embeddings peform better in link prediction in both GCN and LNE models when underlying graphs are intransitive. Error bars reflect multiple iterations with different random seeds as in \citet{chami2019hyperbolic}.}
    \label{GCN_fig}
\end{figure}

In Figure \ref{GCN_fig} we plot the test set AUC for these models. In graphs which had high intransitivity in Table \ref{global_AR} we see a large gain from using $AR$ as opposed to Euclidean embeddings in both GCN and LNE models.

\newpage

\bibliography{imcl_ar_2023.bbl}
\bibliographystyle{icml2022}

\newpage
\appendix
\onecolumn
\section{Appendix}
\begin{proof}[Proof of Proposition \ref{thm:euc_exists}]
Let $G$ be an arbitrary graph. We will construct an embedding that represents it.

Let $N$ be the number of nodes, let $D$ be a vector in $\mathbb{R}^{N}$. Let $M_{D}$ be an $N \times N$ matrix where $m_{ij} = e_{ij}$ for $i \neq j$ and $D$ is the matrix diagonal. If $M_{D}$ is positive semi-definite then there exists a factorization $M_{D} = VV'.$ 

Row vectors of $V$ are embeddings for each node that represent the graph $G$. We can see now that any $D$ which makes $M_{D}$ positive semi-definite gives us these required embeddings. Such a $D$ always exists since we can always construct one by taking $D$ arbitrary, computing the largest negative eigenvalue $\lambda_{min}$ and then the matrix $M_{D + \lambda_{min}}$ will be positive semi-definite. Clearly the family of embeddings that represent the graph can be put into $1-1$ correspondence with the set of matrix diagonals that make $M_{D}$ positive semi-definite.
\end{proof}

\begin{proof}[Proof of Proposition \ref{thm:no_euclid}]
In essence, we want to determine the rank, that is, determine the dimension of the nullspace of a matrix 
\[
   M = \left( \begin{array}{cc} A & R \\ R & B \end{array} \right)
\]
where $A$ is an $n \times n$ diagonal matrix with coefficients $a_1\dots a_n > 0$, $B$ is an $n \times n$ diagonal matrix with coefficients $b_2\dots b_n > 0$, and $R$ is an $n \times n$ matrix of all $1$.  Let's solve!
\[
    M\times\left(\begin{array}{c} 
      \mathbf{u} \\ \mathbf{v} \end{array} \right) = 0 
      \quad \Longleftrightarrow \quad 
      \forall i  \left\{\begin{array}{ll} 
         a_i u_i + \sum_j v_j = 0 \\
         b_i v_i + \sum_j u_j = 0
    \end{array}\right.
\]
Since $a_i>0$ and $b_i>0$, this implies
\begin{equation}
\label{eq:uvfromsums}
    \forall i \quad
       u_i = - \frac{1}{a_i} \sum_j v_j  \quad
       v_i = - \frac{1}{b_i} \sum_j u_j
\end{equation}
If we were free to set $\sum_j v_j$ and $\sum_j u_j$ as we please,
the two equations~\eqref{eq:uvfromsums} would describe a 2-dimensional space. Therefore the nullspace of $M$ has dimension at most 2. However we can also use the first of these equations to write
\begin{equation}
\label{eq:ufromv}
   \sum_i u_i = - \sum_i \frac{1}{a_i} \sum_j v_j
\end{equation}
Therefore our nullspace has dimension at most 1.
But we can continue and use the second equations from~\eqref{eq:uvfromsums} to replace $v_j$ above:
\[
   \sum_i u_i = -
   \left(\sum_i \frac{1}{a_i}\right)\left(\sum_j v_
]\right) = \left(\sum_i \frac{1}{a_i}\right) \left( \sum_j \frac{1}{b_j}\right) \sum_k u_k
\]
Therefore, if $r = \left(\sum_i \frac{1}{a_i}\right) \left( \sum_j \frac{1}{b_j}\right) \neq 1$, then we must have $\sum_i u_i=\sum_j v_j = 0$ which means that $u_i=v_j=0$: the matrix is nonsingular.
On the other hand, if $r=1$, then I can choose $\sum_j v_j$ equal to any non zero value, deduce $\sum_j u_i$ using~\eqref{eq:ufromv}, compute $u_i$ and $v_i$ using~\eqref{eq:uvfromsums}, and verify that we have described a one-dimensional nullspace.

In conclusion: if $r = \left(\sum_i \frac{1}{a_i}\right) \left( \sum_j \frac{1}{b_j}\right) \neq 1$, the matrix has full rank. If $r=1$ the matrix has rank $2n-1$.  This is the case, for instance, when $a_i=b_j=n$.
\end{proof}

\begin{proof}[Proof of Proposition \ref{thm:nuc_norm}]
By Lemma 6 of \cite{mazumder2010spectral} we know that for a matrix $X$ we can write $$||X||_{*} = \min_{UV = X} \dfrac{1}{2} ( ||U||^2_{F} + ||V||^2_{F}||).$$ With the minimum being attained at the factor decomposition $X = UV$.

Let $M_{D}$ be the minimum nuclear norm solution. By construction of our matrix $M^*$ it has the factor decomposition $U = [A, R]$ and $V = [A, -R]$ where $[\cdot]$ denotes column-wise concatenation.

Substituting the definition of the factors gets $$|| M_{D} ||_{*} = \dfrac{1}{2} (2\sum_{ij} a_{ij}^2 + 2\sum_{ij} r_{ij}^2)$$ which simplifies to $$|| M_{D} ||_{*} = || A ||^2_{F} + || R ||^2_{F}.$$ 

Thus the $A,R$ obtained from $M_{D}$ are in the set of minimum norm $A,R$ decompositions. 
\end{proof}

\begin{proof}[Proof of Proposition \ref{thm:hype_equiv}]
The Lorenz model works as follows: we take a vector in $\mathbb{R}^{d+1}$ written as $(x_0, x_1, \dots, x_d)$ and define the inner product $$\mathcal{L} (x, y) = -x_0 y_0 + \sum_{i = 1}^{d} x_i y_i.$$ This is clearly the pseudo-Euclidean inner product with $R$ dimension $1$. However, while pseudo-Euclidean space is not a metric space, we can take the manifold defined by $$\mathcal{H}^d = \lbrace x \in \mathbb{R}^{d+1} \mid \mathcal{L} (x,x) = -1, x_0 > 0 \rbrace$$ and endow it with the metric $$d(x, y) = \text{arccosh} (-\mathcal{L} (x, y)).$$ This defined manifold and metric is diffeomorphic to the Poincare ball. Given a set of vectors on the Poincare ball, we can take the inverse of this diffeomorphism on these vectors to get their Lorenz counterparts. Then since $d(x,y)$ here is a monotone transformation of $\mathcal{L} (x,y)$, we have that if $d(x,y) > d(x,z)$ then $\mathcal{L} (x,y) < \mathcal{L} (x,z).$ Thus we have constructed an AR embedding that order-represents the graph.
\end{proof}

\subsection{Expanded Ingredient Substitution List}
In the main text we reported the top $R$ neighbor for each focal ingredient to save space. Here we report the top $3$ neighbors per each focal ingredient. 

\begin{table}{r}
\centering
\begin{tiny}
\begin{tabular}{p{1.5cm} p{1.8cm} p{1cm}}
  \hline
target & substitute & R score \\ 
  \hline
baking mix & bisquick & 0.78 \\ 
  baking mix & biscuit mix & 0.73 \\ 
  baking mix & bisquick mix & 0.70 \\ 
  baking powder & baking soda & 0.85 \\ 
  baking powder & whole wheat flour & 0.51 \\ 
  baking powder & all-purpose flour & 0.42 \\ 
  beer & apple juice & 0.40 \\ 
  beer & mango & 0.39 \\ 
  beer & corn oil & 0.39 \\ 
  brown sugar & sugar & 0.74 \\ 
  brown sugar & honey & 0.69 \\ 
  brown sugar & light brown sugar & 0.68 \\ 
  buttermilk & skim milk & 0.52 \\ 
  buttermilk & soymilk & 0.48 \\ 
  buttermilk & chickpea & 0.39 \\ 
  chicken broth & chicken stock & 0.85 \\ 
  chicken broth & vegetable broth & 0.63 \\ 
  chicken broth & vegetable stock & 0.61 \\ 
  lemon & fresh lemon juice & 0.76 \\ 
  lemon & lemon, juice of & 0.71 \\ 
  lemon & lemon juice & 0.66 \\ 
  onion & red onion & 0.71 \\ 
  onion & scallion & 0.71 \\ 
  onion & yellow onion & 0.68 \\ 
  orange juice & honey & 0.61 \\ 
  orange juice & orange & 0.50 \\ 
  orange juice & lemon & 0.47 \\ 
  parmesan cheese & mozzarella & 0.64 \\ 
  parmesan cheese & cheddar & 0.62 \\ 
  parmesan cheese & olive oil & 0.53 \\ 
  parsley & fresh parsley & 0.93 \\ 
  parsley & flat leaf parsley & 0.65 \\ 
  parsley & dried parsley & 0.59 \\ 
  pecan & walnut & 0.85 \\ 
  pecan & nut & 0.75 \\ 
  pecan & sliced almond & 0.65 \\ 
  red wine & dry red wine & 0.79 \\ 
  red wine & dry white wine & 0.68 \\ 
  red wine & white wine & 0.61 \\ 
  unsalted butter & butter & 0.74 \\ 
  unsalted butter & margarine & 0.67 \\ 
  unsalted butter & heavy cream & 0.48 \\ 
  unswtd chocolate & unswtd choc square & 0.83 \\ 
  unswtd chocolate & baking cocoa & 0.74 \\ 
  unswtd chocolate & unswtd cocoa & 0.71 \\ 
  vegetable oil & oil & 0.96 \\ 
  vegetable oil & canola oil & 0.88 \\ 
  vegetable oil & olive oil & 0.67 \\ 
  vinegar & cider vinegar & 0.89 \\ 
  vinegar & white vinegar & 0.87 \\ 
  vinegar & apple cider vinegar & 0.82 \\ 
  yogurt & plain yogurt & 0.73 \\ 
  yogurt & greek yogurt & 0.70 \\ 
  yogurt & vanilla yogurt & 0.53 \\ 
   \hline
\end{tabular}
\caption{Substitutes for various focal ingredients found by looking at cosine similarity neighbors in the $R$ component.}\label{ingsub}
\end{tiny}
\end{table} 

\subsection{Experiment 5 Supplement}\label{appendix_exp5}
\begin{figure}[h!]
    \centering
    \includegraphics[scale=.7]{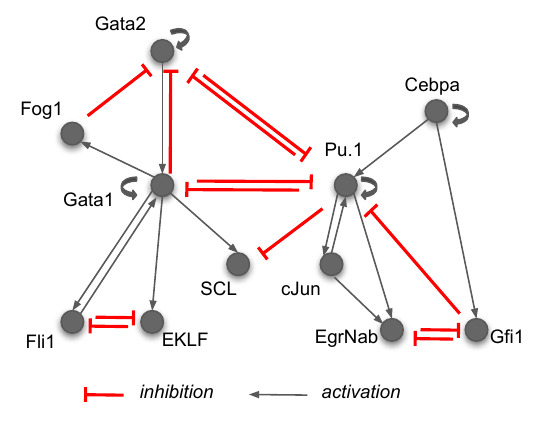} \\
    \includegraphics[scale=.45]{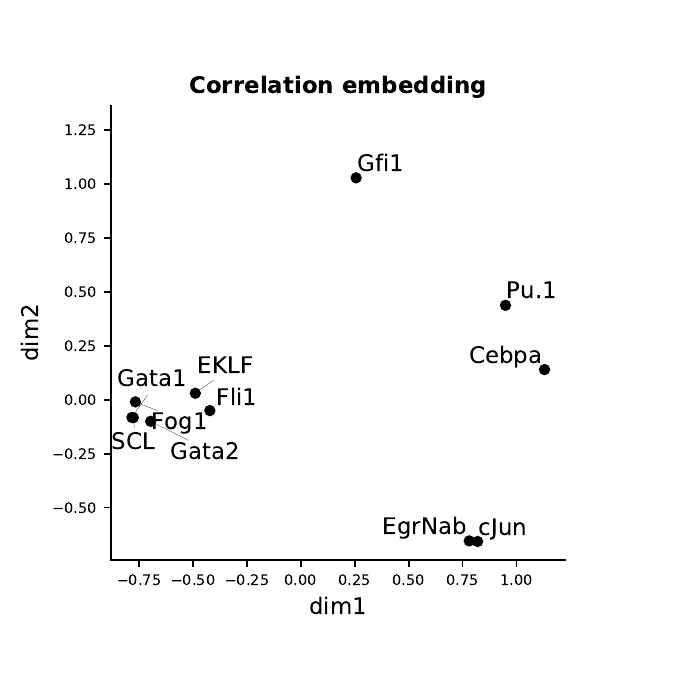}
    \includegraphics[scale=.45]{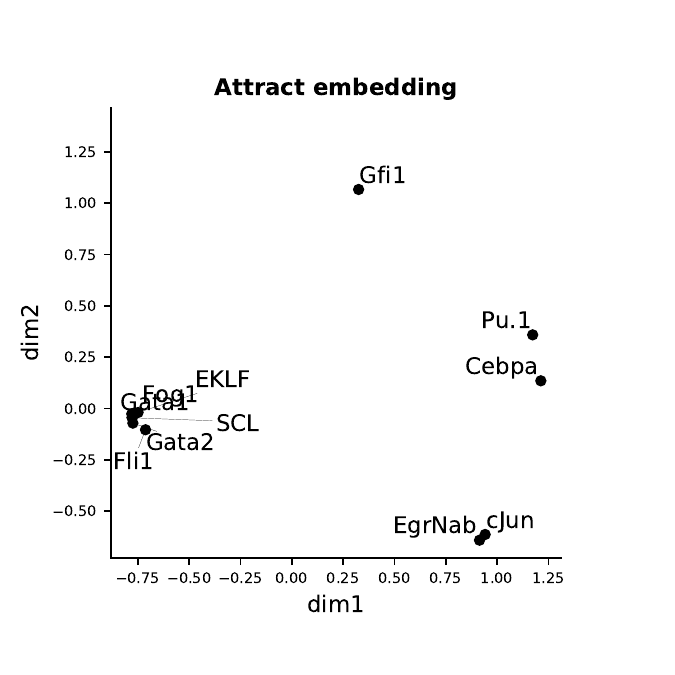}
    \includegraphics[scale=.45]{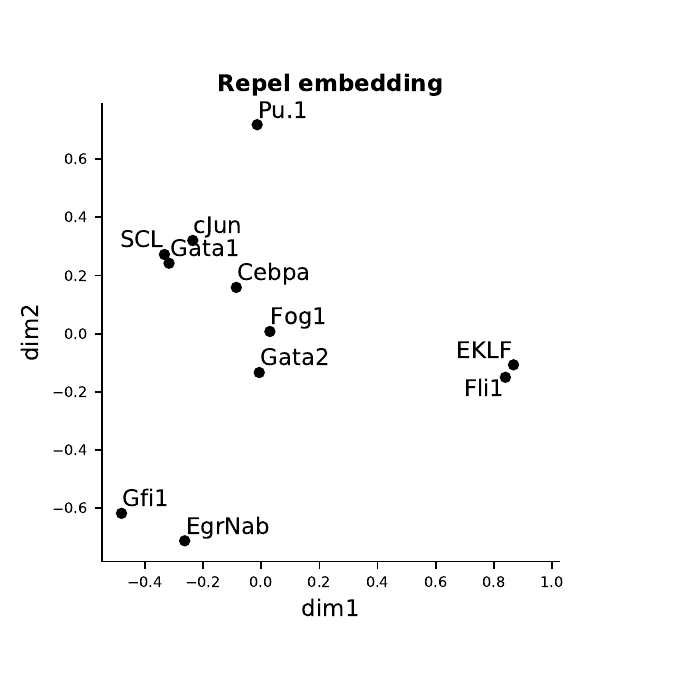}
    \caption{First panel shows the true directed network generating the symmetric co-occurrence patterns we observe with red indicating inhibition and black denoting activation. Other panels show a kernel PCA of the correlation matrix dot product mode as well as the $A$ and $R$ components respectively. While the inhibition/activation structure is well preserved in the $AR$ embeddings, it is not nearly as clear in the standard `attract only' decomposition.}
    \label{fig:krum}
\end{figure}

We show the governing equations of the \citet{krumsiek2011hierarchical} model in \ref{fig:krum}. The original network is represented by boolean rules, which we translated into a system of ODEs to allow us to sample from the model. 

In addition to the analysis in the main text, we use Kernel PCA to visualize the $2$ dimensional projection of the PSD factorization of the Spearman correlation matrix of transcription factors (panel B) compared to the projections of the $A$ (panel C) and $R$ (panel D) components of the AR decomposition.

We see that mutual inhibitors have the highest $R$-similarity scores (and so are close together in the Kernel PCA representation). One way inhibitions have lower scores, which is not surprising, because the inhibitions doesn’t happen immediately and at some points of time both transcription factors can still be observed together. A similar story is obvious in the $A$-similarities showing mutual and one-way activations. However, looking at the embedding of the correlations, such relationships are not obvious. This is partially driven by the fact that in the correlations it is hard to differentiate between two items which have similar contexts but do not appear together (i.e. inhibitors) from items which have low correlation because they are on very different pathways.

\subsection{Experiment 6 Supplement}\label{appendix_exp6}
The architecture of the GCN works as follows: let $X$ be the feature vectors of nodes stacked, a single layer GCN is written as $R(\hat{A} X W^0)$ where $\hat{A}$ is the normalized adjacency matrix and $R$ is some non-linearity. The GCN takes the node features, maps them into a hidden space by the learned $W^0$, takes neighbor averages, and passes them through a non-linearity. This outputs an embedding vector for each node. The graph convolution process can be repeated on this vector again if desired, still outputting one vector per node.

When training the link prediction models of dimension $d$ we always use $\frac{d}{2}$ dimensions for $A$ and $R$ in the LNE-AR/GCN-AR cases. Because of this we found it is important to have different regularization rates for each of the subspaces with the full regularizer being$\lambda_A \sum_i ||a_i|| + \lambda_{R} \sum_i ||r_i||$. An alternative is to keep a single $L2$ regularization weight over all parameters but fix a dimension $d$ and sweep $k$ so that $d-k$ dimensions are $A$ and $k$ are $R$.

We use an $80/10/10$ train/validation/test split. We vary hyperparameters of $d \in \lbrace 12, 24, 48, 96 \rbrace$ and regularization rates $\lambda_i \in \lbrace 1e-7, 1e-6, 1e-5, 1e-4, 1e-3, 1e-2, 1e-1 \rbrace$ for all models. For GCNs we also vary the number of convolution layers (1 or 2) as well as dropout $\in \lbrace 0, .1 \rbrace$, though we did not find dropout or convolution beyond a single layer to be useful in our datasets.

\end{document}